\documentclass[prd,showpacs,showkeys,nofootinbib,preprint]{revtex4-1}
\usepackage{amsmath,amsfonts,amssymb,color}
\usepackage{bm}% bold math
\usepackage{pgfplots}
\usepackage[colorlinks=true,urlcolor=blue,linkcolor=blue,citecolor=blue]{hyperref}
\begin{document}

\title{Conservation laws in gauge gravity theory}

\author{Yuri N. Obukhov}

\affiliation{Russian Academy of Sciences, Nuclear Safety Institute, 
B.Tulskaya 52, 115191 Moscow, Russia
\email{obukhov@ibrae.ac.ru}}

\begin{abstract}
We analyse the conservation laws in the gauge gravity theory which are derived for the general class of gravitational models with the action invariant under the local Poincar\'e and the diffeomorphism group. The consistent Noether-Lagrange formalism is developed, revealing the important role of the auxiliary Goldstone and Stueckelberg fields, with the help of which we construct the composite gauge fields that have a clear geometrical and physical meaning.
\end{abstract}

\maketitle

%%%%%%%%%%%%%%%%%%%%%%%%%%%%%%%%%%%%%%%%%%%%%%%%%%%%%%%%%%%%%%%%%%%%
\section{Introduction}	
%%%%%%%%%%%%%%%%%%%%%%%%%%%%%%%%%%%%%%%%%%%%%%%%%%%%%%%%%%%%%%%%%%%%

Currently, we are witnessing a revival of interest to the gauge theory of gravity \cite{Blagojevic,Reader,MAG,Mielke,selected,overview,Ponomarev}, with a special focus on the Poincar\'e gauge approach and the teleparallel gravity \cite{Aldrovandi,Cho,MH,yno:2003,Obukhov:2002,Koivisto1,Koivisto2}. The latter can be considered as a particular case of the former, if we recall that the 4-parameter group of translations $T_4$ is a subgroup of the 10-parameter Poincar\'e group $G\!=\!T_4\!\rtimes\!SO(1,3)$, the semi-direct product of translations and the Lorentz group. During the recent Tartu conferences, an interesting discussion \cite{Fontanini,Delliou,Huguet1,Huguet2,Pereira} has been started about the structure of the teleparallel gravity, its geometrical interpretation, and the role of the fundamental variables of this theory, namely of the coframe (or the tetrad) and the local Lorentz connection. The present paper contributes to this discussion.

We analyse the conservation laws in the Poincar\'e gauge theory, which are here derived in the framework of the consistent Noether-Lagrange formulation, and demonstrate that these results actually help to clarify the status of the fundamental gauge gravitational variables. Our discussion is local, and the subtle issues of the rigorous fibre bundle picture are put aside here. An interested reader can find the corresponding details in \cite{Reader,Mielke,Ponomarev}. 

It is worthwhile to remind that our basic notation and conventions are those of \cite{MAG}, in particular, Greek indices, $\alpha, \beta,\dots{} = 0, \dots, 3$, denote the anholonomic components. The Minkowski metric reads $g_{\alpha\beta} = {\rm diag}(+1,-1,-1,-1)$.

%%%%%%%%%%%%%%%%%%%%%%%%%%%%%%%%%%%%%%%%%%%%%%%%%%%%%%%%%%%%%%%%%%%%
\section{Gauge gravitational field}
%%%%%%%%%%%%%%%%%%%%%%%%%%%%%%%%%%%%%%%%%%%%%%%%%%%%%%%%%%%%%%%%%%%%

We can consider the affine space $A_4$ at every point of the spacetime manifold as an extension of the tangent space \cite{KN}, on which the action of the Poincar\'e group is defined in a natural way. The Poincar\'e gauge fields then arise as the translational and the Lorentz parts of the {\it generalised affine connection}, see \cite{Reader,Mielke,Ponomarev},
\begin{equation}
\omega^{\alpha},\qquad \omega_{\alpha}{}^{\beta}.\label{PGfields}
\end{equation}
These are 1-forms on the spacetime, taking values in the Lie algebras of the group of translations and the Lorentz group, respectively. Under the action of an arbitrary Poincar\'e group element
\begin{equation}
\left(b^{\alpha},\ L_{\alpha}{}^{\beta}\right),\qquad b^{\alpha}(x)\in T_4,
\ L_{\alpha}{}^{\beta}(x)\in SO(1,3),\label{PG}
\end{equation}
the gauge fields (\ref{PGfields}) have the following transformation laws 
\begin{eqnarray}
\omega^{\prime\alpha} &=& L_{\beta}{}^{\alpha}\left[\omega^{\beta} - d(L^{-1}_{\gamma}{}^{\beta}b^{\gamma})
- \omega_{\delta}{}^{\beta}L^{-1}_{\gamma}{}^{\delta}b^{\gamma}\right],\label{PGtrans2}\\
\omega^{\prime}_{\alpha}{}^{\beta} &=& L_{\mu}{}^{\beta}\omega_{\nu}{}^{\mu}L^{-1}_{\alpha}{}^{\nu}+ 
L_{\gamma}{}^{\beta}dL^{-1}_{\alpha}{}^{\gamma}.\label{PGtrans1}
\end{eqnarray}
In addition to the gauge field potentials, we also introduce a {\it Goldstone type} scalar (i.e., 0-form) field which arises as the section of the affine tangent bundle, 
\begin{equation}
\phi^{\alpha},\label{phi}
\end{equation}
with the transformation law
\begin{equation}
\phi^{\prime\alpha} = L_{\beta}{}^{\alpha}\phi^{\beta} + b^{\alpha}.\label{phiT}
\end{equation}
This variable can be viewed as defining an origin of affine spaces at every point of the spacetime.

The structure equations for the generalised affine connection give rise to the translational and the Lorentz gauge field strengths
\begin{eqnarray}
\Theta^{\alpha} &=& % D\omega^{\alpha}=
d\omega^{\alpha} +\omega_{\beta}{}^{\alpha}\wedge\omega^{\beta},\label{PGstr1}\\
\Omega_{\alpha}{}^{\beta} &=& d\omega_{\alpha}{}^{\beta} + \omega_{\gamma}{}^{\beta}\wedge\omega_{\alpha}{}^{\gamma}.
\label{PGstr2}
\end{eqnarray}
It is straightforward to derive the transformation laws of these objects,
\begin{equation}
\Omega^{\prime}_{\alpha}{}^{\beta} = L_{\delta}{}^{\beta}\Omega_{\gamma}{}^{\delta}L^{-1}_{\alpha}{}^{\gamma},\quad
\Theta^{\prime\alpha} = L_{\beta}{}^{\alpha}\Theta^{\beta} -\omega^{\prime}_{\beta}{}^{\alpha}b^{\beta}.\label{PGstrT}
\end{equation}
Since the Goldstone field transforms covariantly (\ref{phiT}) under the Lorentz subgroup, one can define the relevant covariant derivative with the help of the Lorentz gauge field $\omega_{\alpha}{}^{\beta}$ which has the proper transformation law (\ref{PGtrans1}),
\begin{equation}
D\phi^{\alpha} = d\phi^{\alpha} + \omega_{\beta}{}^{\alpha}\wedge\phi^{\beta}.\label{Dphi}
\end{equation}

%%%%%%%%%%%%%%%%%%%%%%%%%%%%%%%%%%%%%%%%%%%%%%%%%%%%%%%%%%%%%%%%%%%%
\section{Noether-Lagrange formalism}
%%%%%%%%%%%%%%%%%%%%%%%%%%%%%%%%%%%%%%%%%%%%%%%%%%%%%%%%%%%%%%%%%%%%

Now let us develop the Noether-Lagrange machinery for the Poincar\'e gauge theory. We assume that the pure gravitational Lagrangian,
\begin{equation}
L = L(\omega^{\alpha}, d\omega^{\alpha}, \omega_{\alpha}{}^{\beta}, d\omega_{\alpha}{}^{\beta}, 
\phi^{\alpha}, d\phi^{\alpha}), \label{L}
\end{equation}
is a scalar $4$-form invariant under the local Poincar\'e group, and the gravitational action (derived as an integral of $L$ over the spacetime manifold) is also diffeomorphism invariant. At first let us study the consequences of the Poincar\'e gauge invariance. 

From the invariance of $L$ under the local Lorentz group one can, as usual \cite{MAG}, conclude that the Lagrangian has actually the form
\begin{equation}
L=L(\omega^{\alpha}, \Theta^{\alpha}, \Omega_{\alpha}{}^{\beta}, \phi^{\alpha}, D\phi^{\alpha}). \label{L1}
\end{equation}
We need to evaluate the explicit variation
\begin{equation}
\delta  L=\delta \omega^{\alpha}\wedge {\frac {\partial L}{\partial\omega^{\alpha}}} +
\delta \Theta^{\alpha}\wedge{\frac {\partial L}{\partial\Theta^{\alpha}}} +
\delta \Omega_{\beta}{}^{\alpha}\wedge{\frac {\partial L}{\partial\Omega_{\beta}{}^{\alpha}}} +
\delta \phi^{\alpha}{\frac {\partial L}{\partial\phi^{\alpha}}} +
\delta (D\phi^{\alpha})\wedge{\frac {\partial L}{\partial D\phi^{\alpha}}}.\label{varL}
\end{equation}
Using the definitions (\ref{PGstr1}), (\ref{PGstr2}), (\ref{Dphi}), one can recast the total variation into
\begin{eqnarray}
\delta  L &=& \delta \omega^{\alpha}\wedge\Sigma_{\alpha} + \delta \omega_{\beta}{}^{\alpha}\wedge
\Delta_{\alpha}{}^{\beta} + \delta \phi^{\alpha}U_{\alpha}\nonumber\\ && +\,d\left[\delta \omega^{\alpha}\wedge
{\frac {\partial L}{\partial\Theta^{\alpha}}} + \delta \omega_{\beta}{}^{\alpha}\wedge{\frac
{\partial L}{\partial\Omega_{\beta}{}^{\alpha}}} + \delta \phi^{\alpha}{\frac {\partial L}
{\partial D\phi^{\alpha}}}\right],\label{varL1}
\end{eqnarray}
where the following notation is used:
\begin{eqnarray}
\Sigma_{\alpha} &:=& {\frac {\partial L}{\partial\omega^{\alpha}}} + 
D{\frac {\partial L}{\partial\Theta^{\alpha}}},\label{Si}\\
\Delta_{\alpha}{}^{\beta} &:=& \omega^{[\beta}\wedge {\frac {\partial L}{\partial\Theta^{\alpha]}}} +
D{\frac {\partial L}{\partial\Omega_{\beta}{}^{\alpha}}} +
\phi^{[\beta}{\frac {\partial L}{\partial D\phi^{\alpha]}}},\label{De}\\
U_{\alpha} &:=& {\frac {\partial L}{\partial\phi^{\alpha}}} - 
D{\frac {\partial L}{\partial D\phi^{\alpha}}}.\label{U}
\end{eqnarray}

%%%%%%%%%%%%%%%%%%%%%%%%%%%%%%%%%%%%%%%%%%%%%%%%%%%%%%%%%%%%%%%%%%%%
\subsection{Lorentz symmetry}
%%%%%%%%%%%%%%%%%%%%%%%%%%%%%%%%%%%%%%%%%%%%%%%%%%%%%%%%%%%%%%%%%%%%

Nothing prevents us to study the action of the Lorentz and translational subgroups separately. The infinitesimal Lorentz transformations read
\begin{equation}
\delta \omega^{\alpha}=\varepsilon_{\beta}{}^{\alpha}\omega^{\beta},\qquad
\delta \omega_{\alpha}{}^{\beta}=-D\varepsilon_{\alpha}{}^{\beta},\qquad
\delta \phi^{\alpha}=\varepsilon_{\beta}{}^{\alpha}\phi^{\beta},\label{Lor}
\end{equation}
with the antisymmetric parameter functions $\varepsilon_{\beta\alpha} = -\varepsilon_{\alpha\beta}$. 

Substituting this into (\ref{varL1}), one finds that the boundary expression under the exterior differential is zero in view of (\ref{De}), whereas by making use of the arbitrariness of $\varepsilon^{\alpha\beta}$ one obtains the identity
\begin{equation}
\omega_{[\beta}\wedge\Sigma_{\alpha]} + D\Delta_{\alpha\beta} + \phi_{[\beta}U_{\alpha]}=0.\label{consL}
\end{equation}
This looks like the standard Noether identity for the Poincar\'e gauge theory in the presence of the ``matter" field $\phi^{\alpha}$. However, one should notice the principal difference: the translational gauge field $\omega^{\alpha}$ is not a coframe 1-form, and hence the 3-form $\Sigma_{\alpha}$ cannot be viewed as the ordinary canonical energy-momentum current. 

%%%%%%%%%%%%%%%%%%%%%%%%%%%%%%%%%%%%%%%%%%%%%%%%%%%%%%%%%%%%%%%%%%%%
\subsection{Translational symmetry}
%%%%%%%%%%%%%%%%%%%%%%%%%%%%%%%%%%%%%%%%%%%%%%%%%%%%%%%%%%%%%%%%%%%%

Let us turn now to the consequences of the translational invariance. The infinitesimal translations read
\begin{equation}
\delta \omega^{\alpha} = -\,D\varepsilon^{\alpha},\qquad \delta \omega_{\alpha}{}^{\beta}=0,\qquad
\delta \phi^{\alpha} = \varepsilon^{\alpha},\label{trans}
\end{equation}
for arbitrary parameter functions $\varepsilon^{\alpha}$. 

Substituting these into (\ref{varL1}), and reorganizing the terms with derivatives, one finds
\begin{equation}
\delta  L=\varepsilon^{\alpha}\left[D\Sigma_{\alpha} + U_{\alpha}\right] + 
d\left[\,\varepsilon^{\alpha}\!\left(-\,\Sigma_{\alpha} + D{\frac {\partial L}{\partial\Theta^{\alpha}}} +
{\frac {\partial L}{\partial D\phi^{\alpha}}}\right)\right],\label{varL2}
\end{equation}
and hence the independence and arbitrariness of $\varepsilon$ and $d\varepsilon$ yields a pair of identities:
\begin{eqnarray}
D\Sigma_{\alpha} + U_{\alpha} =0,\label{DS}\\
\Sigma_{\alpha} = D{\frac {\partial L}{\partial\Theta^{\alpha}}} +
{\frac {\partial L}{\partial D\phi^{\alpha}}}.\label{Sig}
\end{eqnarray}
These identities play an important role in the further construction of the Poincar\'e gauge theory. In fact, (\ref{DS})-(\ref{Sig}) {\it suggest the  effective elimination of the Goldstone field} $\phi^{\alpha}$ {\it and introduce the coframes as the basic dynamical objects}. Indeed, substitution of (\ref{Sig}) in (\ref{DS})
yields, recalling (\ref{U}),
\begin{equation}
{\frac {\partial L}{\partial\phi^{\alpha}}} = 
\Omega_{\alpha}{}^{\beta}\wedge {\frac {\partial L}{\partial\Theta^{\beta}}},\label{dLdp}
\end{equation}
while directly comparing (\ref{Sig}) with the definition (\ref{Si}), one finds
\begin{equation}\label{dLdDp}
{\frac {\partial L}{\partial D\phi^{\alpha}}} = {\frac {\partial L}{\partial \omega^{\alpha}}}.
\end{equation}

%%%%%%%%%%%%%%%%%%%%%%%%%%%%%%%%%%%%%%%%%%%%%%%%%%%%%%%%%%%%%%%%%%%%
\section{Composite gauge fields}
%%%%%%%%%%%%%%%%%%%%%%%%%%%%%%%%%%%%%%%%%%%%%%%%%%%%%%%%%%%%%%%%%%%%

It is worthwhile to notice that both relations (\ref{dLdp}) and (\ref{dLdDp}) are strong identities which must hold for any Poincar\'e invariant Lagrangian irrespectively of the validity of the field equations. These relations underlie the above mentioned possibility of the effective elimination of $\phi^\alpha$ from the theory. On a formal level one can simply replace everywhere all derivatives of the Lagrangian with respect to $\phi^\alpha$ and $d\phi^\alpha$ by the right hand sides of (\ref{dLdp}) and (\ref{dLdDp}). On a deeper level, the identity (\ref{dLdDp}) tells us that the two variables $\omega^{\alpha}$ and $D\phi^{\alpha}$ can enter the Poincar\'e gauge invariant Lagrangian only in combination $\omega^{\alpha} + D\phi^{\alpha}$. Analogously, (\ref{dLdp}) demands that $\phi^{\alpha}$ and $\Theta^{\alpha}$ can appear in $L$ only in combination $\Theta^{\alpha} + \Omega_{\beta}{}^{\alpha}\phi^{\beta}$. Having noticed this, let us introduce the {\it effective fields}
\begin{eqnarray}
\vartheta^{\alpha} &:=& \omega^{\alpha} + D\phi^{\alpha},\label{cof}\\
T^{\alpha} &:=& \Theta^{\alpha} + \Omega_{\beta}{}^{\alpha}\phi^{\beta}.\label{Tor}
\end{eqnarray}
One can easily verify that the two new objects are related by the differential
identity
\begin{equation}
T^{\alpha}=D\vartheta^{\alpha}.\label{Tor1}
\end{equation}
However, the crucial observation is that these composite objects transform covariantly under the action of the Poincar\'e group. Accordingly, we can consistently interpret $\vartheta^{\alpha}$ as the coframe 1-form, and $T^\alpha$ as the torsion 2-form on the spacetime manifold. 

Thus, summarising, we discover that the gravitational Lagrangian invariant under the local Poincar\'e group (\ref{Lor}) and (\ref{trans}) can be recast into the form
\begin{eqnarray}
L &=& L(\omega^{\alpha}, d\omega^{\alpha}, \omega_{\alpha}{}^{\beta}, d\omega_{\alpha}{}^{\beta}, \phi^{\alpha},
d\phi^{\alpha})\nonumber\\
&=& L(\omega^{\alpha}, \Theta^{\alpha}, \Omega_{\alpha}{}^{\beta}, \phi^{\alpha}, D\phi^{\alpha})=
\widetilde{L}(\vartheta^{\alpha}, T^{\alpha}, \Omega_{\alpha}{}^{\beta}),\label{Lt}
\end{eqnarray}
where the first and the second arguments of $\widetilde{L}$ should be treated as composite fields, which depend on the original fundamental Poincar\'e gauge fields through (\ref{cof})-(\ref{Tor}). It is useful to observe the following relations between the partial derivatives:
\begin{eqnarray}
{\frac {\partial L}{\partial\omega^{\alpha}}} = {\frac {\partial L}{\partial D\phi^{\alpha}}}
&=& {\frac {\partial\widetilde{L}}{\partial\vartheta^{\alpha}}},\label{dLdo1}\\
{\frac {\partial L}{\partial\Theta^{\alpha}}} &=&
{\frac {\partial\widetilde{L}}{\partial T^{\alpha}}},\label{dLdTh}\\
{\frac {\partial L}{\partial\omega_{\alpha}{}^{\beta}}} -
\phi^{[\alpha}{\frac {\partial L}{\partial\Theta^{\beta]}}} &=&
{\frac {\partial\widetilde{L}}{\partial\Omega_{\alpha}{}^{\beta}}}.\label{dLdO}
\end{eqnarray}

Substituting (\ref{dLdp}),(\ref{dLdDp}), and (\ref{dLdo1})-(\ref{dLdO}) into the total variation formula (\ref{varL1}), one can now rewrite it as
\begin{equation}
\delta  L = \delta \widetilde{L} = \delta \vartheta^{\alpha}\wedge\widetilde{\Sigma}_{\alpha} + 
\delta \omega_{\beta}{}^{\alpha}\wedge\widetilde{\Delta}_{\alpha}{}^{\beta} + d\left[\delta \vartheta^{\alpha}
\wedge{\frac {\partial\widetilde{L}}{\partial T^{\alpha}}} + \delta \omega_{\beta}{}^{\alpha}\wedge
{\frac {\partial\widetilde{L}}{\partial\Omega_{\beta}{}^{\alpha}}}\right],\label{varL3}
\end{equation}
where 
\begin{eqnarray}
\widetilde{\Sigma}_{\alpha} &=& {\frac {\partial\widetilde{L}}{\partial\vartheta^{\alpha}}}
+ D{\frac {\partial\widetilde{L}}{\partial T^{\alpha}}},\label{Sit}\\
\widetilde{\Delta}_{\alpha}{}^{\beta} &=& \vartheta^{[\beta}\wedge{\frac {\partial\widetilde{L}}{\partial T^{\alpha]}}} +
D{\frac {\partial\widetilde{L}}{\partial\Omega_{\beta}{}^{\alpha}}}.\label{Det}
\end{eqnarray}
These new objects now qualify for the energy-momentum and the spin currents, respectively. Notice that the original (``fundamental'') spin 3-form is related to the effective one by means of the identity
\begin{equation}\label{DD}
\Delta_{\alpha}{}^{\beta} - \Sigma_{[\alpha}\,\phi^{\beta]} = \widetilde{\Delta}_{\alpha}{}^{\beta}.
\end{equation}

%%%%%%%%%%%%%%%%%%%%%%%%%%%%%%%%%%%%%%%%%%%%%%%%%%%%%%%%%%%%%%%%%%%%
\subsection{Diffeomorphism invariance}
%%%%%%%%%%%%%%%%%%%%%%%%%%%%%%%%%%%%%%%%%%%%%%%%%%%%%%%%%%%%%%%%%%%%

To complete the Noether-Lagrange analysis, let us discuss the consequences of the diffeomorphism invariance of the Poincar\'e gauge theory. It should be stressed, that the diffeomorphism symmetry is understood as a motion of the spacetime manifold generated by smooth vector fields, and in this sense it should be clearly distinguished from the translations considered above.

The derivation of conservation laws is well known \cite{MAG}: Starting from the original Lagrangian, its covariant Lie derivative (one can replace the ordinary Lie derivative in view of the local Lorentz invariance) along an arbitrary vector field $\xi$ reads
\begin{eqnarray}
\delta_{\rm diff}L = {\cal L}_{\xi}L &=&
{\cal L}_{\xi}\omega^{\alpha}\wedge{\frac {\partial L}{\partial\omega^{\alpha}}} +
{\cal L}_{\xi}\Theta^{\alpha}\wedge{\frac {\partial L}{\partial\Theta^{\alpha}}} +
{\cal L}_{\xi}\omega_{\beta}{}^{\alpha}\wedge{\frac {\partial L}{\partial\omega_{\beta}{}^{\alpha}}}\nonumber\\
&& +\,{\cal L}_{\xi}\phi^{\alpha}{\frac {\partial L}{\partial\phi^{\alpha}}} +
{\cal L}_{\xi}(D\phi^{\alpha})\wedge{\frac {\partial L}{\partial D\phi^{\alpha}}}.\label{diffL}
\end{eqnarray}
Proceeding along the usual lines (making use of the ``$A+dB=0$'' scheme, see \cite{MAG}), one finds from (\ref{diffL}) the two identities:
\begin{eqnarray}
(\xi\rfloor \omega^{\alpha})D\Sigma_{\alpha} - (\xi\rfloor D\phi^{\alpha})U_{\alpha} && \nonumber\\
-\,(\xi\rfloor \Theta^{\alpha})\wedge\Sigma_{\alpha} - (\xi\rfloor \Omega_{\alpha}{}^{\beta})\wedge
\Delta_{\beta}{}^{\alpha} &=& 0,\label{diffI1}\\
(\xi\rfloor L) - (\xi\rfloor \omega^{\alpha}){\frac {\partial L}{\partial\omega^{\alpha}}}  
- (\xi\rfloor D\phi^{\alpha}){\frac {\partial L}{\partial D\phi^{\alpha}}}  && \nonumber\\
-\,(\xi\rfloor \Theta^{\alpha})\wedge{\frac {\partial L}{\partial\Theta^{\alpha}}} - (\xi\rfloor
\Omega_{\beta}{}^{\alpha})\wedge{\frac {\partial L}{\partial\Omega_{\beta}{}^{\alpha}}} &=& 0.\label{diffI2}
\end{eqnarray}
These are again quite similar to the standard Noether identities.

All together, the set of identities (\ref{consL}), (\ref{dLdp}), (\ref{dLdDp}), (\ref{diffI1}), (\ref{diffI2}) provide a complete description of the invariance properties of the fundamental Poincar\'e gauge theory (under the local Poincar\'e and the diffeomorphism group) in terms of the original variables (\ref{PGfields}) and (\ref{phi}). However, analogously to the effective form of the total variation (\ref{varL2}), one can rewrite the Noether identities in terms of the effective composite fields (\ref{cof}) and (\ref{Tor}). Inserting the definitions (\ref{cof})-(\ref{Tor}) into the translational identities (\ref{dLdp}) and (\ref{dLdDp}) we find identical zero, while inserting into the Lorentz (\ref{consL}) and diffeomorphism (\ref{diffI1}), (\ref{diffI2}) Noether identities, we find
\begin{equation}
\vartheta^{[\beta}\wedge\widetilde{\Sigma}_{\alpha]} + D\widetilde{\Delta}_{\alpha}{}^{\beta} = 0,\label{consL1}
\end{equation}
and, respectively,
\begin{eqnarray}
(\xi\rfloor\vartheta^{\alpha})D\widetilde{\Sigma}_{\alpha} -
(\xi\rfloor T^{\alpha})\wedge\widetilde{\Sigma}_{\alpha} - 
(\xi\rfloor \Omega_{\alpha}{}^{\beta})\wedge\widetilde{\Delta}_{\beta}{}^{\alpha} &=& 0,\label{consT}\\
(\xi\rfloor\widetilde{L}) - (\xi\rfloor\vartheta^{\alpha})\widetilde{\Sigma}_{\alpha}
+ (\xi\rfloor\vartheta^{\alpha}){\frac {\partial\widetilde{L}}{\partial T^{\alpha}}}  
- (\xi\rfloor T^{\alpha})\wedge{\frac {\partial\widetilde{L}}{\partial T^{\alpha}}} - 
(\xi\rfloor\Omega_{\beta}{}^{\alpha})\wedge{\frac {\partial\widetilde{L}}
{\partial\Omega_{\beta}{}^{\alpha}}} &=& 0.\label{Sig2}
\end{eqnarray}
Specifying from an arbitrary vector field $\xi$ to the vector frame $e_\alpha$ dual to the coframe $\vartheta^{\alpha}$, we recover the standard energy-momentum conservation law in the Poincar\'e gauge gravity theory \cite{Reader,MAG,Mielke,Ponomarev}. 

Our analysis of the symmetry and the conservation laws apparently revealed an important role played by the Goldstone field $\phi^\alpha$. In this relation, a naturally interesting question arises: what is the structure of the Poincar\'e gauge Lagrangian when the Goldstone field $\phi^{\alpha}$ is absent? An immediate consequence of the translational invariance then is, in accordance with (\ref{dLdDp}), the independence of the gravitational Lagrangian on the translational gauge field,
\begin{equation}
{\frac {\partial L}{\partial\omega^{\alpha}}} = 0,\label{dLdo}
\end{equation}
while (\ref{diffI2}) then yields (in four dimensions)
\begin{equation}
L = {\frac 12}\left(\Theta^{\alpha}\wedge{\frac {\partial L}{\partial\Theta^{\alpha}}} + 
\Omega_{\beta}{}^{\alpha}\wedge{\frac {\partial L}{\partial\Omega_{\beta}{}^{\alpha}}}\right).\label{LL}
\end{equation}
This rules out the linear Lagrangian of the Hilbert-Einstein type, however, the purely quadratic models are allowed. In greater detail this issue of theory's structure for this case will be studied elsewhere. 

%%%%%%%%%%%%%%%%%%%%%%%%%%%%%%%%%%%%%%%%%%%%%%%%%%%%%%%%%%%%%%%%%%%%
\section{Teleparallelism as a translational gauge theory}
%%%%%%%%%%%%%%%%%%%%%%%%%%%%%%%%%%%%%%%%%%%%%%%%%%%%%%%%%%%%%%%%%%%%

Let us now consider the gauge theory of the group of translations which formally arises when we reduce the Poincar\'e symmetry (\ref{PG}) to $T_4$ that acts on the local affine space $A_4$ by shifting its points by $b^\alpha$. Then the set of the gauge fields (\ref{PGfields}) reduces to the translational part $\omega^\alpha$ with the transformation law (\ref{PGtrans2}) reduced to
\begin{equation}
\omega^\prime{}^\alpha = \omega^\alpha - db^\alpha.\label{transT}
\end{equation}
The corresponding translational gauge field strength (\ref{PGstr1}) obviously reads $\Theta^\alpha = d\omega^\alpha$.  In a similar way, the transformation law (\ref{phiT}) of the Goldstone field (\ref{phi}) is simplified to a shift
\begin{equation}
\phi^{\prime\alpha} = \phi^{\alpha} + b^{\alpha}.\label{phiTT}
\end{equation}
Repeating the Noether-Lagrange analysis above, we then end up with a new set of composite fields
\begin{eqnarray}
\vartheta^{\alpha} &=& \omega^{\alpha} + d\phi^{\alpha},\label{cof1}\\
T^{\alpha} &=& \Theta^{\alpha} = d\omega^\alpha,\label{Tor2}
\end{eqnarray}
replacing the coframe (\ref{cof}) and the torsion (\ref{Tor}).

Thereby, we naturally recover the standard translational gauge theory \cite{Aldrovandi,Cho}, in which the Lorentz symmetry (and hence the local Lorentz connection) is absent on the fundamental level. However, we can bring the Lorentz connection in by introducing an additional variable $u_\beta{}^\alpha(x) \in SO(1, 3)$ as a local group element. By definition, the action of the local Lorentz group on this variable reads as a usual group multiplication
\begin{equation}
u^{\prime}_\beta{}^{\alpha} = L_{\gamma}{}^{\alpha}u_{\beta}{}^\gamma.\label{uT}
\end{equation}
An extended gravitational Lagrangian of the translational gauge model then reads 
\begin{equation}
L = L(\omega^{\alpha}, d\omega^{\alpha}, u_{\alpha}{}^{\beta}, du_{\alpha}{}^{\beta}, 
\phi^{\alpha}, d\phi^{\alpha}), \label{LT}
\end{equation}
and we can develop the Noether-Lagrange machinery for this model along the lines above. Assuming the invariance of this model under the local translations (\ref{transT}), (\ref{phiTT}) and the local Lorentz transformations (\ref{uT}), it is straightforward to derive the corresponding conservation laws which show that the model (\ref{LT}) is consistently recast into an {\it effective Poincar\'e gauge theory} 
\begin{equation}
L(\omega^{\alpha}, d\omega^{\alpha}, u_{\alpha}{}^{\beta}, du_{\alpha}{}^{\beta}, \phi^{\alpha}, d\phi^{\alpha}) =
\widehat{L}(\widehat{\vartheta}{}^{\alpha}, \widehat{T}{}^{\alpha}),\label{LTT}
\end{equation}
where the composite fields are introduced as follows:
\begin{eqnarray}
\widehat{\vartheta}{}^{\alpha} &=& u_{\beta}{}^{\alpha}\bigl(\omega^{\beta}
+ \widehat{D}\widehat{\phi}{}^{\beta}\bigr),\label{cof2}\\
\widehat{\omega}{}_{\alpha}{}^{\beta} &=& u_{\gamma}{}^{\beta}d\,u^{-1}_{\alpha}{}^{\gamma}.\label{lor2}
\end{eqnarray}
Here $\widehat{\phi}{}^\alpha := u_{\beta}{}^{\alpha}\phi^{\beta}$, and the covariant derivative $\widehat{D}$ is determined by the local Lorentz connection (\ref{lor2}). Obviously, the corresponding curvature vanishes
\begin{equation}\label{nocur}
\widehat{\Omega}{}_{\alpha}{}^{\beta} = d\widehat{\omega}{}_{\alpha}{}^{\beta}
+ \widehat{\omega}{}_{\gamma}{}^{\beta}\wedge\widehat{\omega}{}_{\alpha}{}^{\gamma} \equiv 0,
\end{equation}
and, therefore, this Poincar\'e gauge theory is indeed a {\it teleparallelism} theory. The torsion is easily computed:
\begin{equation}\label{torTT}
\widehat{T}{}^{\alpha} = \widehat{D}\widehat{\vartheta}{}^{\alpha} = u_{\beta}{}^{\alpha}\Omega^{\beta}.
\end{equation}

The resulting theory (\ref{LT})-(\ref{nocur}) is known as the covariant teleparallel gravity \cite{illumi,MarPer} in which the local Lorentz connection is nontrivial. One can check the full covariance of the effective theory, in particular, verify the property (\ref{PGtrans1}) of the connection (\ref{lor2}) induced by the local Lorentz transformation (\ref{uT}). By making use of the latter, one can obviously gauge away the new variable to the value $u_{\alpha}{}^{\beta} = \delta_{\alpha}^{\beta}$, and hence to gauge away the local Lorentz connection, $\widehat{\omega}{}_{\alpha}{}^{\beta} = 0$. In this gauge, the effective model reduces to the original translational gauge theory (\ref{transT})-(\ref{Tor2}). One thus can interpret $u_{\alpha}{}^{\beta}$ as a generalized Stueckelberg field. 

It is worthwhile to mention that the representation (\ref{lor2}) was consistently used by Blixt et al \cite{Blixt} for the thorough analysis of the degrees of freedom issue in the teleparallel gravity theory; see also the related work on the Lorentz symmetry in the teleparallel gravity by Hohmann \cite{Hohmann1,Hohmann2}.

%%%%%%%%%%%%%%%%%%%%%%%%%%%%%%%%%%%%%%%%%%%%%%%%%%%%%%%%%%%%%%%%%%%%
\section{Conclusion}
%%%%%%%%%%%%%%%%%%%%%%%%%%%%%%%%%%%%%%%%%%%%%%%%%%%%%%%%%%%%%%%%%%%%

In this paper we study the conservation laws in the gauge gravity theory which are derived for the general class of models invariant under the local Poincar\'e and diffeomorphism group. The consistent Noether-Lagrange analysis reveals the important role of the $T_4$-valued Goldstone field and the $SO(1,3)$-valued Stueckelberg fields, with the help of which the composite gauge fields are constructed. The latter have a clear geometrical and physical meaning, in contrast to the original gauge gravitational variables.

%%%%%%%%%%%%%%%%%%%%%%%%%%%%%%%%%%%%%%%%%%%%%%%%%%%%%%%%%%%%%%%%%%%%
\section*{Acknowledgments}
%%%%%%%%%%%%%%%%%%%%%%%%%%%%%%%%%%%%%%%%%%%%%%%%%%%%%%%%%%%%%%%%%%%%

This work was partially supported by the Russian Foundation for Basic Research (Grant No. 18-02-40056-mega).

\end{document}